 \documentclass[preprint,review,12pt]{elsarticle2}

\usepackage{amsmath}
\usepackage{amssymb}
\usepackage{caption}
\usepackage{graphicx}
\usepackage{subcaption}
\usepackage{placeins}
\usepackage{color}
\usepackage[normalem]{ulem}
\usepackage{wasysym}
\usepackage{psfrag}
\usepackage{xcolor}

\definecolor{marron}{rgb}{0.78,0.44,0.216}
\definecolor{lightblue}{rgb}{0,1,1}
\definecolor{purple}{rgb}{0.48,0.17,0.96}

\newcommand{\ab}[1]{{\color{black}#1}}

\newcommand{\ft}{\hat} 

\begin{document}

\begin{frontmatter}
\title{Intrinsic aeroacoustic instabilities in  the crosstalk apertures of can-annular combustors}
\author{Audrey Blond\'e} 
\author{Khushboo Pandey}
\author{Bruno Schuermans}
\author{Nicolas Noiray\corref{cor1}}
\address{{CAPS Laboratory, Department of Mechanical and Process Engineering, ETH Zurich, Sonnegstrasse 3, 8092 Zurich, Switzerland}}
\cortext[cor1]{Corresponding author. \textit{E-mail address:} noirayn@ethz.ch}
\begin{abstract}
This paper presents an experimental and numerical study of aeroacoustic instabilities at the interface between neighbouring combustion chambers in modern heavy-duty gas turbines. A simplified laboratory-scale geometry of the gap separating the outlet of these chambers, just upstream of the turbine inlet in can-annular combustor architectures, is considered. It consists of two channels with anechoic and chocked conditions on the upstream and downstream sides respectively. Right before the choked-flow vanes which represent the turbine inlet, a small aperture leads to an aeroacoustic crosstalk between the channels. The dimensions and flow conditions are defined such that relevant Mach, Strouhal and Helmholtz numbers of gas turbines are reproduced. The alignment of the vanes with respect to the crosstalk aperture is  varied. An intense whistling is observed for some conditions.  The oscillation frequency depends on the aperture area and scales with the Strouhal number based on the aperture length. The upstream anechoic  condition in each channel implies that no longitudinal acoustic mode participate to the mechanism of this whistling, which is in agreement with the Strouhal scaling of this intrinsic aeroacoustic instability. It is shown that the geometry of the upstream edge of the aperture is an essential element in the occurrence and intensity of the whistling. Compressible Large Eddy Simulations of the configuration have been performed and remarkably reproduce the whistling phenomenon. A detailed investigation of the numerical results revealed the region of sound production from the shear layer oscillations in the crosstalk aperture. This work contributes to the understanding of  aeroacoustic instabilities at the crosstalk apertures of can-annular combustors. It will help designing combustor-turbine interfaces to suppress them, which is important since the vibrations they induce may be as damaging as the ones from thermoacoustic instabilities.
\end{abstract}


\end{frontmatter}

\section{Introduction}
\label{sec:intro}
\begin{figure} 
    \centering
    \begin{psfrags}
    \psfrag{combustor}[][][1]{\scriptsize \begin{tabular}{@{}c@{}}
Individual\\[-2pt]
can combustors
\end{tabular} }
    \psfrag{vanes}[][][1]{\scriptsize First row of turbine vanes}
    \psfrag{wall}[][][1]{\scriptsize \begin{tabular}{@{}c@{}}
\textcolor{orange}{Combustor}\\[-2pt]
\textcolor{orange}{wall}
\end{tabular}}
\psfrag{XT}[][][1]{\scriptsize \textcolor{red}{Crosstalk}}
\psfrag{sound}[][][1]{\scriptsize \textcolor{marron}{Sound production}}
\psfrag{layer}[][][1]{\scriptsize \textcolor{blue}{Shear layer}}
\psfrag{plane}[][][1]{\scriptsize \textcolor{purple}{Choked flow}}
\psfrag{a}[][][1]{\scriptsize a)}
\psfrag{b}[][][1]{\scriptsize b)}
\psfrag{c}[][][1]{\scriptsize c)}
\psfrag{d}[][][1]{\scriptsize d)}
\includegraphics[width=17cm]{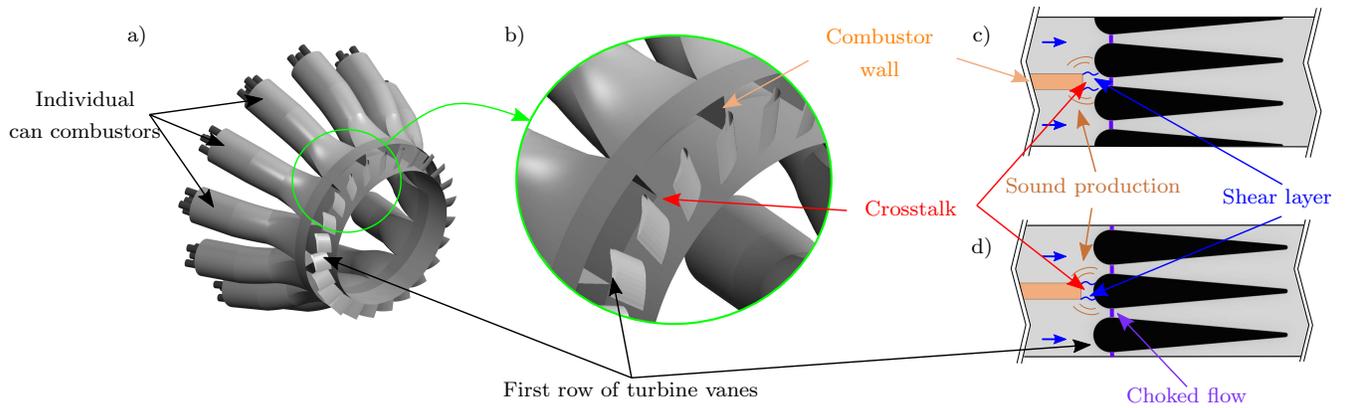}
    \end{psfrags}
    \caption{(a-b) Illustration of a typical can-annular combustor (adapted from \cite{moon2020cross}). (c-d) Generic geometries of the crosstalk area considered in the present study, with aligned and misaligned combustor wall with respect to the leading edge of the simplified turbine vanes. In contrast with the practical gas turbine geometry, the vanes are straight. This simplified design facilitated the development and manufacturing of the experimental configuration, without altering the relevance of the geometry for the study of aeroacoustic instabilities, because the occurence of these instabilities does not  depend on the high Mach region downstream of the choked flow plane.}
    \label{fig:CanComb} 
\end{figure}

To meet the Paris Agreements objectives on CO$_2$ emissions, the global energy system has to be rapidly decarbonised. The production of electricity from wind and solar sources increases, but it is intermittent and must be balanced to satisfy the demand. One option is to convert the electric energy excess in chemical fuels that can be stored and burned on demand in gas turbines  to compensate the negative fluctuations of these renewable sources \cite{gotz2016renewable,glenk2019economics}. Hydrogen can be produced from water electrolysis and it constitutes the basis all more complex gaseous or liquid electrofuels. It can be blended with natural gas to supply modern gas turbines, and the manufacturers are currently working on the development of new gas turbines capable of burning gaseous fuel blends up to pure hydrogen. A challenging aspect of the development of these future fuel-flexible technologies is the problem of thermoacoustic instabilities in combustion chambers \cite{ETN2020}. Thermoacoustic instabilities occur when the feedback mechanisms between the unsteady heat release rate of the flame and
the acoustic field in the combustor lead to acoustic energy production that exceeds acoustic energy dissipation \cite{schuermans2022rayleigh}. \\
The largest gas turbines produce several hundreds of MW and exhibit can-annular combustor architectures. They feature a ring of can-combustors that acoustically communicate with each other before the first stage of the turbine. The small opening between the combustor outlet and the turbine inlet is  referred to as the crosstalk aperture. Figures \ref{fig:CanComb}a and \ref{fig:CanComb}b illustrate a typical can-annular combustor.\\
The thermoacoustic behavior of can-annular combustors has been the topic of several studies from industry: Siemens \cite{bethke2002thermoacoustic,kaufmann20083d,farisco2017thermo}, General Electrics  \cite{venkatesan2019heavy,moon2020cross} or Ansaldo Energia  \cite{ghirardo2019thermoacoustics}. The presence of the crosstalk aperture between neighbouring cans constitutes a challenge for transferring knowledge on thermoacoustics from single can tests to full can-annular combustor. Indeed, this acoustic coupling can lead to instabilities involving global acoustic modes that do not exist in single can-combustors. The coupling strength depends on the acoustic compactness of the crosstalk aperture. In the weakly coupled can-annular system the eigenvalues are clustered in the complex plane, and it was shown in \cite{Pedergnana20242192} that the  eigenmodes continuously transform into the azimuthal mode of an annular combustor when the aperture length is increased. As a side note, the presence of these aperture between the cans is not only a challenge for the prediction of self-sustained acoustic oscillations, but also for the developing the cooling concept of the interface between the combustor and the turbine and its aerothermal performance \cite{luque2015new}.\\
This acoustic crosstalk between cans significantly influence the  thermoacoustic instabilities of the can-annular combustor, and it has motivated many studies in the last few years, e.g.  \cite{jegal2019mutual,moon2019combustion,fournier2021low,fournier2022interplay,von2021non}. Moon \textit{et al.} \cite{moon2020cross} investigate the different types of acoustic self-oscillations (push-pull or push-push) that a four-can combustor can experience when the individual combustors communicate via transversal cross tubes. It was also shown that the modal dynamics in a similar can-annular system strongly differs between odd and even numbers of cans  \cite{moon2023modal} and depends on can-to-can asymmetries in terms of  flame response to acoustic perturbations \cite{moon2021influence,jegal2021influence}. Yoon \textit{et al.} \cite{yoon2022thermoacoustics} derive a model for predicting the linear stability for systems with crosstalks in the plenum and  in the combustion chamber. Jegal \textit{et al.} \cite{jegal2021influence} explicitly investigate the effect of asymmetry of flame acoustic responses on the thermoacoustic stability of a combustor equipped with four cans. By changing the swirl number in the cans, they show that this asymmetry changes the nature of the modes between the individual combustors and can be employed as a passive control parameter to tune the thermoacoustic stability. Guan \textit{et al.} \cite{guan2022effect} reinterpret these modal patterns using chimera states. Pedergnana \textit{et al.} \cite{pedergnana2022steady} 
propose an analytical study on the nonlinear and stochastic thermoacoustic dynamics of can-annular combustors exhibiting weak can-to-can coupling. They offer explanations for the intermittent energy transfer between Bloch waves observed in acoustic pressure spectrograms observed of real-world
gas turbines \cite{ghirardo2019thermoacoustics}. 
 The clusters of eigenvalues of can-annular combustors are further studied analytically and experimentally by Humbert \textit{et al.} \cite{humbert2023acoustics}. In their numerical investigation, Farisco \textit{et al.} \cite{farisco2017thermo} show the effect of crosstalk opening on the acoustic transmission coefficients between two cans. They quantify the reflection and transmission coefficient of the outlet of the combustion chambers in presence of crosstalk opening. With their research on the prediction of thermoacoustic stability in a can-annular combustors, Haeringer \textit{et al.} \cite{haeringer2021strategy,haeringer2019time} present a strategy to tune the downstream acoustic boundaries in single can combustors  to mimic  thermoacoustic dynamics of full can-annular systems. \\
Recently, Pedergnana and Noiray \cite{pedergnana2022coupling} investigated 
the aeroacoustic dynamics of the crosstalk area and its potential key role in the onset of self-sustained acoustic oscillations. Their linear stability analysis deals with the case of acoustically compact apertures. It is based on an analytical description of the  aeroacoustic response of the apertures to acoustic forcing, which builds upon the seminal work of Howe on sound production and dissipation of apertures subject to grazing flows \cite{howe1980dissipation}. The aeroacoustic response of the crosstalk aperture was also considered with a linear formulation of the effective crosstalk impedance by Orchini \textit{et al.} \cite{orchini2022effective,orchini2022reduced}. It is important to stress that the prediction of limit cycle amplitudes would necessitate an amplitude dependent effective impedance and that the associated nonlinear dependence is expected to be similar to the one in the case of a side aperture in a pipe flow \cite{pedergnana2021modeling,boujo2018saturation}. \\
The crosstalks between cans are apertures with two-sided low-Mach grazing flows composed of the hot products of the combustion chambers, which is a flow configuration resembling  to the one investigated in numerous works on grazing flow apertures.\\
 However, the typical geometry of apertures found in can-annular combustors differs from the canonical examples considered by Howe and his coworkers \cite{howe1998acoustics,howe1997influence,grace1998influence} and from the generic geometries experimentally studied in \cite{kooijman2008acoustical} or \cite{bourquard2021whistling}, because  it  includes the presence of choked flow immediately downstream of the grazing flow aperture at the turbine inlet.  The corresponding sonic flow region reflects sound waves (see e.g. \cite{weilenmann2021experiments}), which results in significantly different acoustic boundary conditions around the grazing flow aperture compared to the configurations explored in \cite{howe1998acoustics,howe1997influence,grace1998influence,kooijman2008acoustical,bourquard2021whistling}.\\
 Moreover, the typical geometry and compressible flow found at the turbine inlet of practical can-annular combustors (thin shear layers developing from thin wall apertures) also differ from the ones investigated in recent experimental studies on can-annular system's thermoacoustics, e.g. \cite{moon2023modal}, which use crosstalk apertures having very different aerodynamic and aeroacoustic properties, such as transverse tubes.\\
In this context, no experimental results were published so far in the literature on the aeroacoustics of crosstalk apertures with relevant geometry upstream of choked flow vanes. The present work aims at filling this knowledge gap which is crucial for the development of predictive tools of thermoacoustic and aeroacoustic instabilities in real gas turbines. Indeed, the low-order models of the above-mentioned recent theoretical  studies \cite{pedergnana2022coupling,orchini2022reduced} showed  that the shear layer aeroacoustics can have a significant impact on the self-sustained acoustic oscillations 
by either amplifying or suppressing them and therefore be equally important as the thermoacoustic feedback of the flames. \\
Here, we investigate experimentally and numerically the aeroacoustic coupling for two different vanes arrangements, in order to explore the effect of misalignment between the aperture edge and the closest vane. The motivation for studying the influence of this misalignment is the fact that usually, the number of vanes in the first row of the turbine is not a multiple of the number of cans in a gas turbine which results is several possible outlet geometries. Our unique and modular experimental setup, which  includes upstream twin high-pressure anechoic chambers, allows us to  unravel the existence of self-sustained aeroacoustic instabilities at the crosstalk aperture in absence of flames. This is an important discovery as this non-reactive flow mechanism could be the actual root cause of self-sustained acoustic oscillations observed in real gas turbines which may wrongly be attributed to the acoustic flame interactions. The  crosstalk aperture has been designed to mimic the outlet of two neighbouring cans and the first row of turbine vanes. This canonical geometry is shown in Figs. \ref{fig:CanComb}c and \ref{fig:CanComb}d.  A description of the experimental setup is given in Section \ref{sec:expSetup}. The different configurations of vanes and crosstalk geometries are detailed. Experimental results are discussed in Section \ref{sec:expe}. Compressible large-eddy simulations (LES) of the self-sustained aeroacoustic oscillations were also performed in this work. They are compared to the experimental results in Section \ref{sec:LES} and enable the identification of the acoustic energy production regions in the crosstalk aperture. 

\section{Experimental setup}
\label{sec:expSetup}
\subsection{Test rig}
\begin{figure} 
	    \centering
         \begin{psfrags}
     \psfrag{inlet}[][][1]{\scriptsize \begin{tabular}{@{}c@{}}
Continuous air supply\\[-6pt]
from compressor
\end{tabular}}
     \psfrag{cut}[][][1]{\scriptsize \begin{tabular}{@{}c@{}}
Upstream pressure vessel\\[-6pt]
(cut view) with acoustic\\[-6pt]
absorbing foam (yellow)
\end{tabular}}
     \psfrag{horns}[][][1]{\scriptsize \begin{tabular}{@{}c@{}}
Upstream pressure vessel\\[-6pt]
(cut view) with acoustic\\[-6pt]
absorbing foam (yellow)
\end{tabular}}
\psfrag{horns}[][][1]{\scriptsize \begin{tabular}{@{}c@{}}
Internal catenoidal \\[-6pt]
horns
\end{tabular}}
\psfrag{vessel}[][][1]{\scriptsize \begin{tabular}{@{}c@{}}
Downstream pressure vessel \\[-6pt] (assembled view)
\end{tabular}}
\psfrag{outlet}[][][1]{\scriptsize Exhaust line}
\psfrag{speak}[][][1]{\scriptsize Loudspeaker }
\psfrag{a}[][][1]{\scriptsize a) Overview }
\psfrag{b}[][][1]{\scriptsize b) Zoom-in A-A}
\psfrag{c}[][][1]{\scriptsize c) B-B: Misaligned Vanes}
\psfrag{f}[][][1]{\scriptsize d) B-B: Aligned Vanes}
\psfrag{d}[][][1]{\scriptsize e) Splitting Plate Trailing Edge}
\psfrag{x}[][][1]{\scriptsize A}
\psfrag{y}[][][1]{\scriptsize B}
\psfrag{cold}[][][1]{\scriptsize Cold}
\psfrag{air}[][][1]{\scriptsize air}
\psfrag{mic}[][][1]{\scriptsize Microphones}
\psfrag{micW}[][][1]{\scriptsize \textcolor{red}{Microphone (spectra)}}
\psfrag{xa}[][][1.2]{\scriptsize $x_1$}
\psfrag{xb}[][][1.2]{\scriptsize $x_2$}
\psfrag{xc}[][][1.2]{\scriptsize $x_3$}
\psfrag{xd}[][][1.2]{\scriptsize $x_4$}
\psfrag{xe}[][][1.2]{\scriptsize $x_5$}
\psfrag{xf}[][][1.2]{\scriptsize $x_6$}
\psfrag{xg}[][][1.2]{\scriptsize $x_7$}
\psfrag{xh}[][][1.2]{\scriptsize $x_8$}
\psfrag{x0}[][][1.2]{\scriptsize $x=0$}
\psfrag{0}[][][1]{\scriptsize 0}
\psfrag{25}[][][1]{\scriptsize 25}
\psfrag{83}[][][1]{\scriptsize 83.6}
\psfrag{3}[][][1]{\scriptsize 3.8}
\psfrag{57}[][][1]{\scriptsize 57}
\psfrag{7}[][][1]{\scriptsize 7}
\psfrag{r1}[][][1]{\scriptsize R 3.5}
\psfrag{r2}[][][1]{\scriptsize R 1}
\psfrag{r3}[][][1]{\scriptsize $r_v$ 7.6}
\psfrag{18}[][][1]{\scriptsize 1.88}
\psfrag{e}[][][1]{\scriptsize $d$}
\psfrag{rect}[][][1]{\scriptsize rect.}
\psfrag{round}[][][1]{\scriptsize round}
\psfrag{upper}[][][1]{\scriptsize Upper channel}
\psfrag{lower}[][][1]{\scriptsize Lower channel}
\psfrag{0}[][][1]{\scriptsize 0}
\psfrag{0p2}[][][1]{\scriptsize }
\psfrag{0p4}[][][1]{\scriptsize }
\psfrag{0p6}[][][1]{\scriptsize }
\psfrag{0p8}[][][1]{\scriptsize }
\psfrag{1p0}[][][1]{\scriptsize 1}
\psfrag{500}[][][1]{\scriptsize }
\psfrag{1000}[][][1]{\scriptsize 1000}
\psfrag{1500}[][][1]{\scriptsize }
\psfrag{2000}[][][1]{\scriptsize 2000}
\psfrag{2500}[][][1]{\scriptsize }
\psfrag{3000}[][][1]{\scriptsize 3000}
\psfrag{fq}[][][1]{\scriptsize Freq. [Hz]}
\psfrag{abs}[][][1]{\scriptsize $|R|$}
\psfrag{inlet1}[][][1]{\scriptsize Inlet}
\psfrag{outlet1}[][][1]{\scriptsize Outlet}
\psfrag{vanes}[][][1]{\scriptsize Vanes}
\psfrag{XT}[][][1]{\scriptsize crosstalk aperture}
\psfrag{plate}[][][1]{\scriptsize \textcolor{orange}{Splitting Plate}}
\includegraphics[width=17cm]{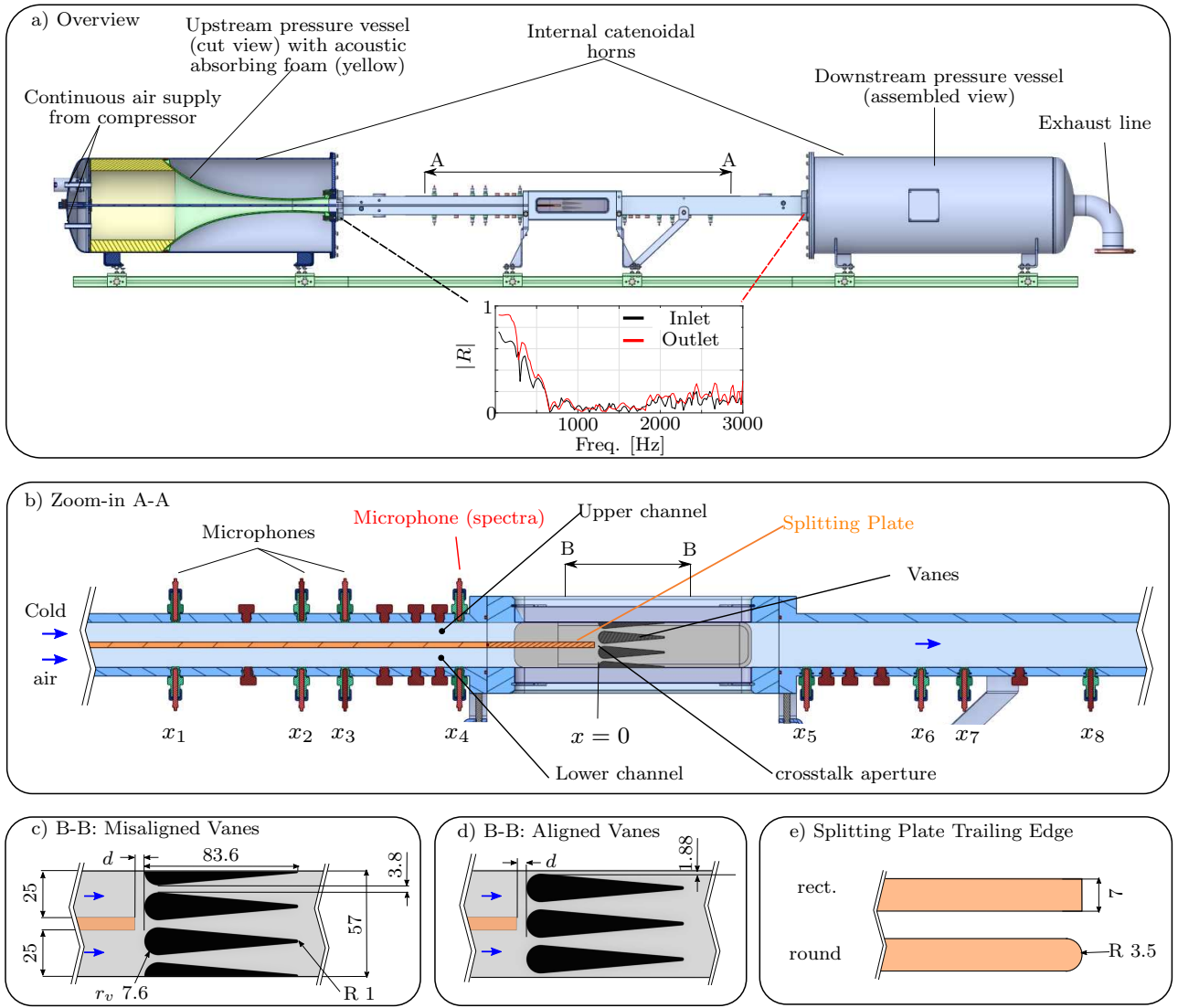}
\end{psfrags}
\caption{a) Overview of the test rig, continuously supplied by an air compressor. b) Zoomed-in cut view of the test section. Microphone locations are specified in Table \ref{tab:mics}. Different geometries can be inserted in the B-B section. c) Misaligned vanes configuration: the splitting plate trailing edge and the vane leading edge are misaligned. d) Aligned vanes. e) Two types of trailing edge of the splitting plate. Dimensions are given in millimeters.
}
\label{fig:setup} 
\end{figure}

The experiments have been conducted in a non-reactive setup shown in Figure \ref{fig:setup}a. 
The experimental setup is composed of two parallel wind channels with a square cross-section followed by a section in which different vanes and crosstalk geometries can be mounted. Figures \ref{fig:setup}c and \ref{fig:setup}d show the different geometries that can be mounted in the section B-B in Fig. \ref{fig:setup}b.
The two channels can be continuously supplied with compressed air at their inlet. They are referred as the upper and lower channels. Each channel features a cross section of 25$\times$25 mm$^2$. The wall separating the two wind channels is referred as the splitting plate (orange in Fig. \ref{fig:setup}b) and its length and trailing edge shape can be adjusted. It represents the combustor wall in a real gas turbine. Its thickness equals 7 mm. After the test section (cut B-B in Fig. \ref{fig:setup}b), the air flows in a rectangular cross-section of area 25$\times$57 mm$^2$. 
At the inlets and outlet, the test rig features two catenoidal horns with acoustic absorbing foams. 

The test rig is operated with a total air mass flow rate of 100 g/s split equally between the two channels (i.e. 50 g/s in each channel). The mass flow rates are measured using Bronkhorst mass flow meters. The temperatures and static pressures are recorded throughout the experiments in both upstream channels and downstream of the vanes. \ab{It is worth mentioning that only operating conditions that choke the flow at the vanes are considered here. This is done to mimic real gas turbines conditions. Once the flow is choked, the air mass flow rate, the pressure and the density in the channels do not vary independently. Consequently, changes of the mass flow rate do not significantly change the flow velocity.}
A total of twelve flush-mounted 1/4” microphones (GRAS Type 46BD-FV) are used: four in the upper upstream channel, four in the lower upstream channel and four downstream of the vanes (cf. Fig. \ref{fig:setup}b). Their signals are recorded either to characterize the acoustic reflection of the upstream and downstream terminations or to obtain the acoustic power spectral density (PSD). For the PSD, the microphone highlighted in red in Fig. \ref{fig:setup}b is used. To compute the acoustic spectra, time traces are recorded for 120 s at a sampling rate of 50 kHz. 
The different geometries that can be mounted in the test rig (section B-B) are depicted in Figure \ref{fig:setup}c and \ref{fig:setup}e. They all feature vanes profile that mimic the first row of turbine vanes. The vanes have been designed to match typical  Helmholtz and Strouhal numbers of industrial geometries. Their dimensions in millimeters are depicted in Fig. \ref{fig:setup}c and \ref{fig:setup}d. Two different vane configurations are considered. They differ by the alignment between the splitting plate and the vanes leading edges. In the misaligned vane configuration sketched in Fig. \ref{fig:setup}c,  the splitting plate does not face the leading edge of a vane directly. The crosstalk distance is then defined as the axial distance between the trailing edge of the splitting plate and the vane leading edge. Fig. \ref{fig:setup}c depicts the latter distance $d$. In the aligned vane configuration represented in Fig. \ref{fig:setup}d, the splitting plate trailing edge is aligned with the central vane. Similarly, the crosstalk distance $d$ is shown in Fig. \ref{fig:setup}d. With these two geometries, the two extreme cases of turbine vane alignment at the outlet of a real gas turbine combustor are investigated. Seven distances $d$ are considered in the present study and summarized in Table \ref{tab:Xtalk_d}.
\begin{table}
    \centering
    \begin{tabular}{|c|c|c|c|c|c|c|c|}
    \hline
    Geometry & 1&2&3&4&5&6&7\\
    \hline
    $d$ [mm]  &1.2& 3.7&6.2&7.2&8.7&9.7&11.2\\
    \hline
    \end{tabular}
    \caption{Lengths of the crosstalk aperture. }
    \label{tab:Xtalk_d}
\end{table}
Finally, two types of plates are considered here: the first one features a rectangular trailing edge whereas the second features a round trailing edge (see Fig. \ref{fig:setup}e).

\definecolor{mygreen}{rgb}{0, 0.5, 0}

\subsection{Operating Conditions}
For each operating condition, the critical area is compared to the geometrical area to ensure that we reach choked flow condition in the vanes passages. From isentropic relationships, the mass flow $\dot{m}^*$ to reach choked conditions, i.e. $M =1$, is expressed as: 
\begin{equation}
    \dot{m}^* = \dfrac{A^* P}{\sqrt{T}} \sqrt{\dfrac{\gamma}{R}} \left(1 + \dfrac{\gamma -1}{2}\right)^{-\frac{\gamma +1}{2\left(\gamma -1\right)}} \; ,
\end{equation}
where $P$ is the pressure, $T$ the temperature, $\gamma$ the ratio of specific heats, $R$ the ideal gas constant and $A^*$ the critical area. 
The critical area $A^*$ can then be retrieved from the operating conditions and compared to the geometrical area. In the present paper, the geometrical area corresponds to $A_{geom} = 1.5 A_{vanes}$, where $A_{vanes}$ is the smallest area between two vane profiles. This is because both the upper and lower channels face 1.5 times the flow passage between two vanes. Here, $A_{vanes} = 3.75 \times 25$ mm$^2$. Choked conditions are reached when the ratio $A^*/ A_{geom}$ is equals or higher than 1. As the compressor sucks air from outside environment, the temperature and pressure may vary from day to day. For the measurements conducted with the misaligned vane configuration, the recorded mean pressure and temperature equal 1.49 bar and 300.2 K, respectively. This leads to a ratio $A^*/ A_{geom}$ equal to 1.02. 
For the experiments conducted with the aligned vane configuration, the recorded mean pressure and temperature equal 1.52 bar and 301.2 K, respectively. This leads to a ratio $A^*/ A_{geom}$ equal to 1.01.

\subsection{Characterisation of the Anechoic Boundary Conditions}

To characterize the cut-off frequency of the upstream and downstream acoustic terminations, a loudspeaker (BEYMA SW1600Nd) is mounted on the upper upstream channel and generates pure tone acoustic forcing. The signals recorded by the microphones mounted on the lower channel are employed in the Multi-Microphone Method (MMM) to reconstruct the acoustic reflection coefficient of the upstream termination. Their positions with respect to the reference $x=0$, labelled from $x_1$ to $x_4$ in Fig. \ref{fig:setup}b are given in Table \ref{tab:mics}. The reader is referred to \cite{schuermans99} for more details on this method and to \cite{weilenmann2021experiments}  for its application with the same test rig without the splitting plate. The microphones, numbered from 8 to 12, are used to retrieve the acoustic reflection coefficient of the downstream termination. Their positions, labelled from $x_5$ to $x_8$ in Fig. \ref{fig:setup}b are given in Table \ref{tab:mics}. 
\begin{table} 
    \centering
     \begin{tabular}{|c|c|c|c|c|c|c|c|c|}
    \hline
    $x_i$ & $x_1$ &$x_2$ &$x_3$ &$x_4$ &$x_5$ &$x_6$ &$x_7$ &$x_8$\\
    \hline
    Pos. [mm] &  $-536$ & $-376$ &$-321$ &$-176$  &266 &411 &466 &626 \\   
    \hline
    \end{tabular}
    \caption{Position of the microphones with respect to the reference location. }
    \label{tab:mics}
\end{table}
Figure \ref{fig:setup}a shows the modulus of these reflection coefficients as function of frequency. 
The black and red solid lines present the reflection coefficients of the upstream and downstream horns, respectively. Above 600 Hz,  the terminations can be considered as acoustically anechoic, i.e. most of the acoustic energy travelling towards the terminations is absorbed and not reflected back to the test rig.

\section{Experimental Results}
\label{sec:expe}
\subsection{Misaligned vanes configuration}
\begin{figure} 
\centering

\begin{psfrags}
\psfrag{0}[][][1]{\scriptsize 0}
\psfrag{80}[][][1]{\scriptsize 80}
\psfrag{100}[][][1]{\scriptsize 100}
\psfrag{120}[][][1]{\scriptsize 120}
\psfrag{140}[][][1]{\scriptsize 140}
\psfrag{70}[][][1]{\scriptsize }
\psfrag{90}[][][1]{\scriptsize }
\psfrag{110}[][][1]{\scriptsize }
\psfrag{130}[][][1]{\scriptsize }
\psfrag{150}[][][1]{\scriptsize }
\psfrag{1000}[][][1]{\scriptsize }
\psfrag{2000}[][][1]{\scriptsize 2000}
\psfrag{3000}[][][1]{\scriptsize 3000}
\psfrag{4000}[][][1]{\scriptsize 4000}
\psfrag{5000}[][][1]{\scriptsize 5000}
\psfrag{6000}[][][1]{\scriptsize 6000}
\psfrag{7000}[][][1]{\scriptsize }
\psfrag{8000}[][][1]{\scriptsize 8000}
\psfrag{9000}[][][1]{\scriptsize }
\psfrag{10000}[][][1]{\scriptsize 10000}
\psfrag{1.2}[][][1]{\scriptsize 1.2}
\psfrag{Rectangular Edge}[][][1]{\scriptsize Rect. Edge}
\psfrag{     Round Edge}[][][1]{\scriptsize Round Edge}
\psfrag{PSD}[][][1]{\scriptsize SPL [dB]}
\psfrag{f}[][][1]{\scriptsize Freq. [Hz]}
\psfrag{a}[][][1]{\scriptsize a)}
\psfrag{b}[][][1]{\scriptsize b)}
\psfrag{dx}[][][1]{\scriptsize $d$ [mm]}
\psfrag{d1}[][][1]{\scriptsize }
\psfrag{d2}[][][1]{\scriptsize 3.7}
\psfrag{d3}[][][1]{\scriptsize 6.2}
\psfrag{d4}[][][1]{\scriptsize 7.2}
\psfrag{d5}[][][1]{\scriptsize 8.7}
\psfrag{d6}[][][1]{\scriptsize 9.7}
\psfrag{d7}[][][1]{\scriptsize 11.2}
\psfrag{2}[][][1]{\scriptsize 2}
\psfrag{4}[][][1]{\scriptsize 4}
\psfrag{6}[][][1]{\scriptsize 6}
\psfrag{8}[][][1]{\scriptsize 8}
\psfrag{10}[][][1]{\scriptsize 10}
\psfrag{12}[][][1]{\scriptsize 12}
\psfrag{rect}[][][1]{\scriptsize \textbf{Rect. Edge}}
\psfrag{round}[][][1]{\scriptsize \textbf{Round Edge}}
\psfrag{c}[][][1]{\scriptsize c)}
\psfrag{d}[][][1]{\scriptsize d)}
\psfrag{e}[][][1]{\scriptsize e)}
\psfrag{pdf}[][][1]{\scriptsize PDF}
\psfrag{1000}[][][1]{\scriptsize 1000}
\psfrag{-1000}[][][1]{\scriptsize -1000}
\psfrag{500}[][][1]{\scriptsize 500}
\psfrag{-500}[][][1]{\scriptsize -500}
\psfrag{amp}[][][1]{\scriptsize Amplitude [Pa]}
\psfrag{st1}[][][1]{\scriptsize $\text{St} = 0.62$}
\psfrag{st2}[][][1]{\scriptsize $\text{St} = 0.51$}
\includegraphics[width=17cm]{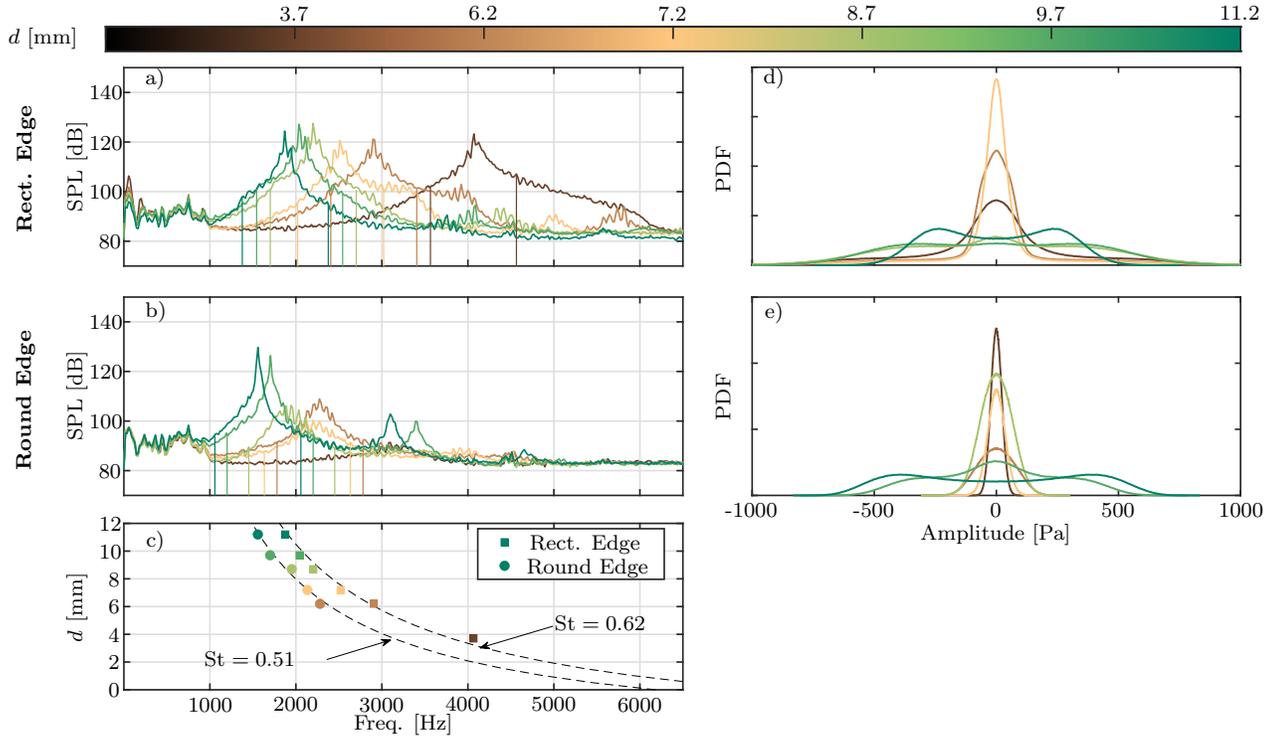}
\end{psfrags}
\caption{Acoustic data recorded for the misaligned vanes configuration, with the rectangular trailing edge (a,d), and with the round edge (b,d). a) and b): Acoustic power spectral density (PSD) in decibels for rectangular and round trailing edge respectively. The colormap  corresponds to various crosstalk opening. c) Frequencies of the dominant spectral peak above 90 dB for rectangular edges (square symbols) and round edges (round symbols). Dashed lines show Strouhal number scaling. d) and e): Probability Density Function (PDF) of the acoustic pressure amplitude recorded by the microphone highlighted in red in Fig. \ref{fig:setup} for rectangular and round trailing edge respectively.} 
\label{fig:mis_all} 
\end{figure}

Experimental results for the misaligned vanes configuration (Fig. \ref{fig:setup}c) are first considered and shown in Figure \ref{fig:mis_all}. 
Data shown in this figure were obtained by post-processing the acoustic signal recorded with the microphone highlighted in red in Fig. \ref{fig:setup}b. Figure \ref{fig:mis_all}a shows the acoustic spectra in decibels for the geometry featuring the rectangular splitting plate. The sound pressure level SPL is defined in decibels as 
$\text{SPL [dB]} = 10 \log_{10} \left( S_{pp}/p_{ref}^2 \right)$, where $S_{pp}$ is the power spectrum of the acoustic pressure and $p_{ref} = 20\,\mu \text{Pa}$ is the reference pressure. 
The color map specifies the crosstalk opening in millimeters and is used in all subsequent figures. 
In Fig. \ref{fig:setup}a, one can see a sharp peak for all crosstalk opening. This is the signature of an intense whistling. Additionally, the comparison of the acoustic spectra denotes that the whistling frequency depends upon the crosstalk opening: they decrease with increasing crosstalk opening distance $d$. The acoustic spectra for the round trailing edge of the splitting plate are plotted in Fig. \ref{fig:mis_all}b for the different crosstalk openings. It is observed from the comparison between Fig. \ref{fig:mis_all}a and \ref{fig:mis_all}b that the geometry of the trailing edge of the splitting plate, and thus of the combustor wall, impacts the occurrence and intensity of the whistling. The crosstalk opening of 3.7 mm does not exhibit any whistling (no peak in the acoustic spectrum). The intermediate aperture lengths of 6.2, 7.2 and 8.7 mm exhibit a broader peak than for the rectangular plates. Moreover, the sound pressure levels associated to these geometries are lower, indicating a weaker sound production with round edge. The whistling frequency is also influenced by the plate geometry, as shown in Fig. \ref{fig:mis_all}c where the frequencies of dominant peaks higher than 90 dB are plotted against $d$. Rectangular and round symbols represent the rectangular and round edge geometries, respectively. The colour coding is the same as in Fig. \ref{fig:mis_all}a and \ref{fig:mis_all}b. First, one can see in Figure~\ref{fig:mis_all}c that the whistling frequency decreases when increasing the aperture length, both for the rectangular and round plates. Second, it highlights the difference in whistling frequencies between rectangular and round edges: the round edge geometries appear to experience whistling at lower frequencies, which can be associated to a different effective gap length. Last, a clear Strouhal scaling of the whistling frequency can be infered from these experiments. The Strouhal number is defined as 
\begin{equation}
    \text{St} = \dfrac{f (d + r_v/2)}{U} \; ,
    \label{eq:Strouhal}
\end{equation}
where $f$, $d$, $r_v$ and $U$ are the frequency, the crosstalk aperture length, the vane leading edge radius and the flow velocity in one channel, respectively. As shown in Fig. \ref{fig:setup}c, $r_v = 7.6$ mm. The dashed black lines in Fig. \ref{fig:mis_all}c represent the Strouhal scaling. The frequencies for the rectangular and round edges scale with a Strouhal number of 0.62 and 0.51, respectively. \\
It is worth noting that all whistling frequencies are above 600 Hz. As illustrated in Section \ref{sec:expSetup}c, the inlet and outlet terminations are acoustically anechoic above 600 Hz. Consequently, the whistling is not due to an aeroacoustic coupling between the shear layer and the rig acoustic modes. As in the case of the Aeolian tone \cite{Etkin195730}, the Strouhal scaling demonstates that the whistling is governed by the flow instability, without feedback mechanism from a longitudinal mode of the system,  and we thus refer  to it as an intrinsic aeroacoustic instability. It is interesting to note that similarly, intrinsic thermoacoustic instabilities \cite{Emmert201575,Silva2023,Gopalakrishnan2023,Dupuy2024}  can develop in combustion chambers equipped with anechoic terminations. \\
To further characterize the whistling phenomenon, the probability density functions (PDF) of the acoustic pressure recorded by the microphone are plotted for the rectangular trailing edge (Fig. \ref{fig:mis_all}d) and for the round one (Fig. \ref{fig:mis_all}e). 
The PDF shape  is directly linked to the linear stability of the aeroacoustic system:  a Gaussian-like distribution corresponds to the sound produced by a linearly stable shear layer mode being stochastically forced by the turbulent fluctuations of the channel flow, and a bimodal distribution corresponds to the case of stochastically forced self-oscillating shear layer mode. This change of the PDF shape is a characteristic feature of randomly forced oscillators exhibiting the dynamics of a Van der Pol oscillator with weak nonlinearity (see e.g. \cite{noiray2017}), when they 
undergo a transition from resonance oscillations to limit cycle oscillations due to a bifurcation parameter change.  Plotting histograms of the band-pass filtered acoustic
data is therefore a good criterion to identify the linear stability of the present aeroacoustic system. For conditions with dominant peak  higher than 90 dB, the acoustic pressure signal is bandpass filtered with a 1000 Hz bandwidth filter centered around the peak. If the peak is lower than 90 dB, the signal is filtered between 1000 Hz and 4000 Hz. The filter bandwidth for each aperture length are represented by the vertical solids lines in Figs. \ref{fig:mis_all}a and \ref{fig:mis_all}b. 
In Figs. \ref{fig:mis_all}d  and \ref{fig:mis_all}e, the change of PDF shape from Gaussian like to bimodal indicate a crossing of the whistling threshold, i.e. a transition from a linearly-stable shear layer mode to a self-oscillating shear layer mode, when increasing the crosstalk aperture length.
\begin{figure} 
\centering
\begin{psfrags}
\psfrag{a}[][][1.2]{\scriptsize a)}
\psfrag{b}[][][1.2]{\scriptsize b)}
\psfrag{amp}[][][1.2]{\scriptsize $\Re(\hat{p}/\hat{p}_{max})$}
\psfrag{phs}[][][1.2]{\scriptsize $\angle \hat{p}$ [rad]}
\psfrag{1}[][][1.2]{\scriptsize 1}
\psfrag{0}[][][1.2]{\scriptsize 0}
\psfrag{-1}[][][1.2]{\scriptsize -1}
\psfrag{-pp}[][][1.2]{\scriptsize $-\pi$}
\psfrag{pp}[][][1.2]{\scriptsize $\pi$}
\psfrag{-0.6}[][][1.2]{\scriptsize -0.6}
\psfrag{-0.5}[][][1.2]{\scriptsize -0.5}
\psfrag{-0.4}[][][1.2]{\scriptsize -0.4}
\psfrag{-0.3}[][][1.2]{\scriptsize -0.3}
\psfrag{-0.2}[][][1.2]{\scriptsize -0.2}
\psfrag{-0.1}[][][1.2]{\scriptsize -0.1}
\psfrag{x}[][][1.2]{\scriptsize Axial Position $x$ [m]}
\psfrag{flow}[][][1.2]{\scriptsize \hspace{-3cm}Flow}
\psfrag{vanes}[][][1.2]{}
\psfrag{upperchanel}[][][1.1]{\scriptsize \,\,Upper Channel}
\psfrag{lowerchannel}[][][1.1]{\scriptsize Lower Channel}
\includegraphics[width=17.5cm]{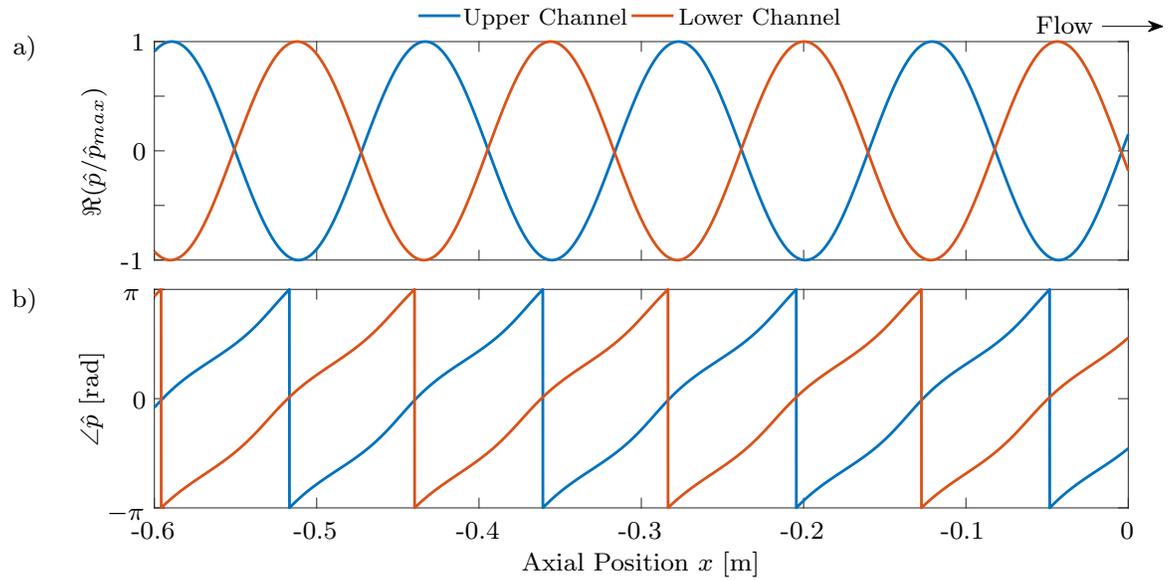}
\end{psfrags}
\caption{Reconstructed acoustic pressure fields for $d=7.2$ mm and the misaligned vanes configuration using the multi-microphone method. Red and blue lines represent the field in the upper and lower channels, respectively. Subfigure a) shows the normalized acoustic pressure amplitude as function of the axial position. Subfigure b) depicts the phase angle in radians of the reconstructed complex pressure field. The vanes location ($x= 0$) and flow direction are labelled on the right-hand side. }
\label{fig:pfield}
\end{figure}\\
\ab{The acoustic pressure field in both channels is reconstructed with the MMM for the misaligned vane configuration and  $d=7.2$ mm and presented in Fig. \ref{fig:pfield}. Similar conclusions can be drawn from other whistling geometries presented in this paper. The averaged complex amplitude of the forward and backward propagating waves $f$ and $g$ are obtained from the 120 s recording. The complex acoustic pressure ansatz $\hat{p}$ is: 
\begin{equation}
   \ft{p}(\omega,x)= \ft{f}(\omega) e^{-i \frac{\omega}{c} \frac{x}{1+\mathrm{M}}} + \ft{g}(\omega) e^{i \frac{\omega}{c} \frac{x}{1-\mathrm{M}}} \; , 
    \label{eq:p_mmm}
\end{equation}
where $\omega$, $c$, $M$ and $x$ are the angular frequency, the speed of sound, the Mach number in the channel and the axial position, respectively. \\
Figure \ref{fig:pfield}a depicts the normalized reconstructed acoustic pressure field at an arbitrary time, corresponding to the real part of the complex pressure field from Eq. (\ref{eq:p_mmm}), $\Re(\hat{p}/\hat{p}_{max})$. The red and blue  lines represent the acoustic pressure field in the upper and lower channels, respectively. The axial location $x$ is given with respect to the reference location $x=0$ indicated in Fig. \ref{fig:setup}b. Figure \ref{fig:pfield}a shows that the pressures in the upper and lower channels are in phase opposition. Figure \ref{fig:pfield}b shows the phase of the complex pressure. One can see that the phases are opposed in the upper and lower channels and that they decrease for decreasing $x$, meaning that there is an upward acoustic propagation from $x=0$ which is in agreement with the aeroacoustic  source located at the crosstalk aperture.}  
\ab{As indicated in Section \ref{sec:expSetup}, once the choked flow condition has been reached at the vanes by increasing the mass flow, the bulk flow velocity in the channels stays constant. Consequently, for a given aperture length $d$, the frequency oscillation will also not change for these mass flows if the Strouhal scaling remains valid. This independence of the whistling frequency on the operating pressure once the flow is choked at the vanes  is presented in the appendix.}

\subsection{Aligned Vanes Configuration}
\begin{figure} 
\centering
\begin{psfrags}
\psfrag{0}[][][1]{\scriptsize 0}
\psfrag{80}[][][1]{\scriptsize 80}
\psfrag{100}[][][1]{\scriptsize 100}
\psfrag{120}[][][1]{\scriptsize 120}
\psfrag{140}[][][1]{\scriptsize 140}
\psfrag{70}[][][1]{\scriptsize }
\psfrag{90}[][][1]{\scriptsize }
\psfrag{110}[][][1]{\scriptsize }
\psfrag{130}[][][1]{\scriptsize }
\psfrag{150}[][][1]{\scriptsize }
\psfrag{1000}[][][1]{\scriptsize }
\psfrag{2000}[][][1]{\scriptsize 2000}
\psfrag{3000}[][][1]{\scriptsize 3000}
\psfrag{4000}[][][1]{\scriptsize 4000}
\psfrag{5000}[][][1]{\scriptsize 5000
}
\psfrag{6000}[][][1]{\scriptsize 6000}
\psfrag{7000}[][][1]{\scriptsize }
\psfrag{8000}[][][1]{\scriptsize 8000}
\psfrag{9000}[][][1]{\scriptsize }
\psfrag{10000}[][][1]{\scriptsize 10000}
\psfrag{1.2}[][][1]{\scriptsize 1.2}
\psfrag{Rectangular Edge}[][][1]{\scriptsize Rect. Edge}
\psfrag{     Round Edge}[][][1]{\scriptsize Round Edge}
\psfrag{PSD}[][][1]{\scriptsize SPL [dB]}
\psfrag{f}[][][1]{\scriptsize Freq. [Hz]}
\psfrag{a}[][][1]{\scriptsize a)}
\psfrag{b}[][][1]{\scriptsize b)}
\psfrag{dx}[][][1]{\scriptsize $d$ [mm]}
\psfrag{d1}[][][1]{\scriptsize 1.2}
\psfrag{d2}[][][1]{\scriptsize 3.7}
\psfrag{d3}[][][1]{\scriptsize 6.2}
\psfrag{d4}[][][1]{\scriptsize 7.2}
\psfrag{d5}[][][1]{\scriptsize 8.7}
\psfrag{d6}[][][1]{\scriptsize 9.7}
\psfrag{d7}[][][1]{\scriptsize 11.2}
\psfrag{2}[][][1]{\scriptsize 2}
\psfrag{4}[][][1]{\scriptsize 4}
\psfrag{6}[][][1]{\scriptsize 6}
\psfrag{8}[][][1]{\scriptsize 8}
\psfrag{10}[][][1]{\scriptsize 10}
\psfrag{12}[][][1]{\scriptsize 12}
\psfrag{rect}[][][1]{\scriptsize \textbf{Rect. Edge}}
\psfrag{round}[][][1]{\scriptsize \textbf{Round Edge}}
\psfrag{c}[][][1]{\scriptsize c)}
\psfrag{d}[][][1]{\scriptsize d)}
\psfrag{e}[][][1]{\scriptsize e)}
\psfrag{pdf}[][][1]{\scriptsize PDF}
\psfrag{1000}[][][1]{\scriptsize 1000}
\psfrag{-1000}[][][1.1]{\scriptsize -1000}
\psfrag{500}[][][1]{\scriptsize 500}
\psfrag{-500}[][][1]{\scriptsize -500}
\psfrag{amp}[][][1.1]{\scriptsize Amplitude [Pa]}
\psfrag{st1}[][][1.1]{\scriptsize $\text{St} = 0.55$}
\psfrag{st2}[][][1.1]{\scriptsize $\text{St} = 0.41$}
\includegraphics[width=17cm]{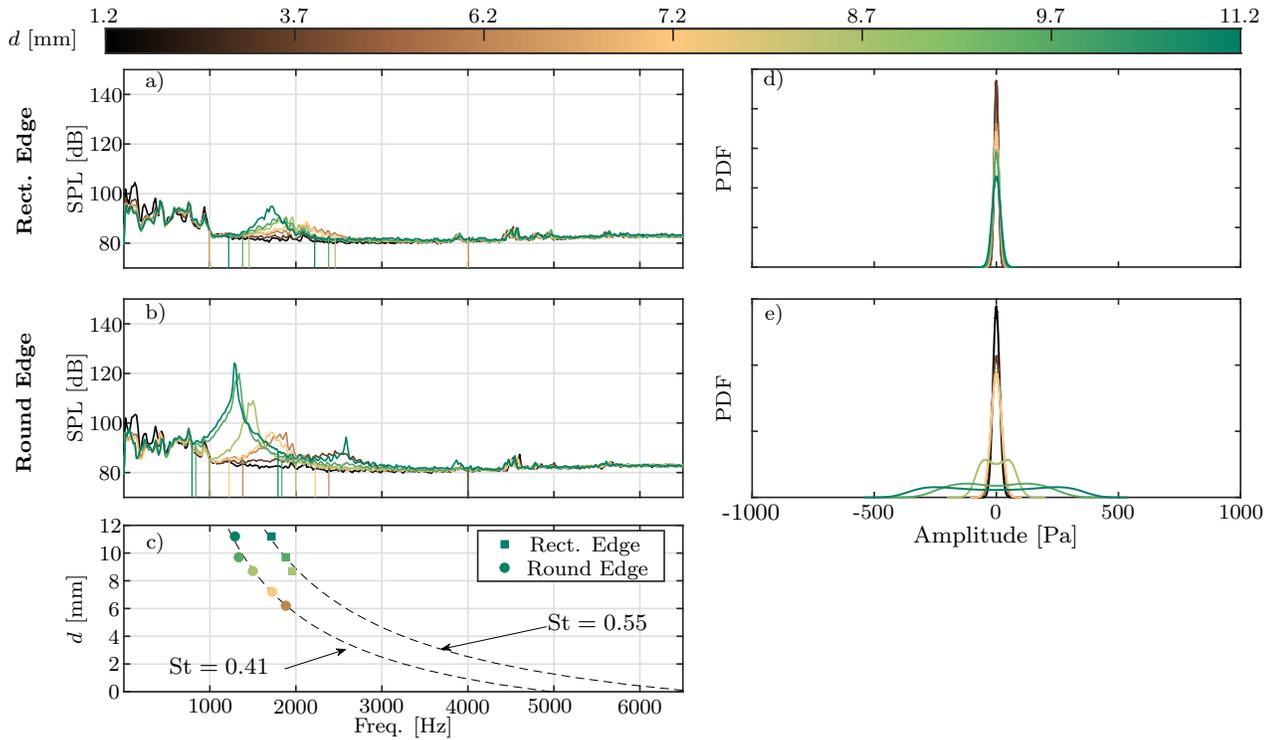}
\end{psfrags}
\caption{Acoustic data recorded for the aligned vane configurations, with the rectangular trailing edge (a,d), and with the round edge (b,d). a) and b): Acoustic power spectral density (PSD) in decibels for rectangular and round trailing edge respectively. The colormap  corresponds to various crosstalk opening. c) Frequencies of the dominant spectral peak above 90 dB for rectangular edges (square symbols) and round edges (round symbols). Dashed lines show Strouhal number scaling. d) and e): Probability Density Function (PDF) of the acoustic pressure amplitude recorded by the microphone highlighted in red in Fig. \ref{fig:setup} for rectangular and round trailing edge respectively. }
\label{fig:alig_all} 
\end{figure}
Experimental results for the aligned vanes configuration (Fig. \ref{fig:setup}d) are considered in this section. 
Figure \ref{fig:alig_all}a shows the acoustic PSD for the rectangular plate and  different crosstalk aperture length $d$. In contrast with the misaligned configuration presented before, no whistling occurs in the aligned vane configuration with a rectangular plate. This major difference between the two types of geometric configurations highlights the potential impact of the alignment of the  turbine vane with the can-to-can apertures on the onset of aeroacoustic instabilities. Similarly, the acoustic PSD for round plate edges are plotted in Fig. \ref{fig:alig_all}b. This configuration exhibits sharp spectral peaks for the crosstalk aperture lengths $d=8.7$, 9.7 and 11.2 mm. Frequencies of the dominant spectral peaks higher than 90 dB are plotted in Fig. \ref{fig:alig_all}c. Similarly to the case of the misaligned vanes, the peak frequencies decrease with increasing $d$. In the aligned vane case, the Strouhal scaling is obtained for $\mathrm{St}= 0.55$ and 0.41, for the rectangular and round trailing edge respectively.\\
Figures \ref{fig:alig_all}d and \ref{fig:alig_all}e show the PDF of the band-pass filtered acoustic pressure. One can conclude from the PDF that the aeroacoustic system is linearly stable for all the crosstalk geometries featuring rectangular plate edge and aligned vanes. However, in the cases of round trailing edge, the change of PDF shape from Gaussian like to bimodal indicate a crossing of the whistling threshold around $d=8$ mm. Indeed, the bimodal distribution indicating a self-oscillating shear layer mode is found for $d=8.7$, 9.7 and 11.2 mm. Finally, it is worth stressing that these results show a clear difference in aeroacoustic behavior depending on the alignment between the vane and the upstream edge of the crosstalk aperture, as well as on its geometry. These differences are expected to have a crucial role on the onset of aeroacoustic instabilities just upstream of the turbine inlet in a heavy duty can-annular gas turbine. 

\section{Compressible Large Eddy Simulations}
\label{sec:LES}
\subsection{Numerical Setup }
\begin{figure} 
\begin{psfrags}
\psfrag{(a)}[][][1.1]{\scriptsize \textbf{(a)}}
\psfrag{(b)}[][][1.1]{\scriptsize \textbf{(b)}}
\psfrag{(d)}[][][1.1]{\scriptsize \textbf{(c)}}
\psfrag{z}[][][1]{\scriptsize z}
\psfrag{x}[][][1]{\scriptsize x}
\psfrag{y}[][][1]{\scriptsize y}
\psfrag{0.1}[][][1]{\scriptsize 0.1}
\psfrag{4.5}[][][1]{\scriptsize 4.5}
\psfrag{-1.5}[][][1]{\scriptsize -1.5}
\psfrag{+1.5}[][][1]{\scriptsize 1.5}
\psfrag{1.5}[][][1]{\scriptsize 1.5}
\psfrag{Ma}[][][1]{\scriptsize Ma}
\psfrag{Omega}[][][1]{\scriptsize $\Omega_{y}$}
\psfrag{mm}[][][1]{\scriptsize Mesh size(mm)}
\psfrag{Inlet 1}[][][1]{\scriptsize Inlet 1}
\psfrag{Inlet 2}[][][1]{\scriptsize Inlet 2}
\psfrag{x105}[][][0.8]{\scriptsize $(\times 10^5)$}
\psfrag{Qcrit ~ 7 x 106}[][][1]{\scriptsize $Q_{crit} \sim 7\times10^6$}
\includegraphics[width=17cm]{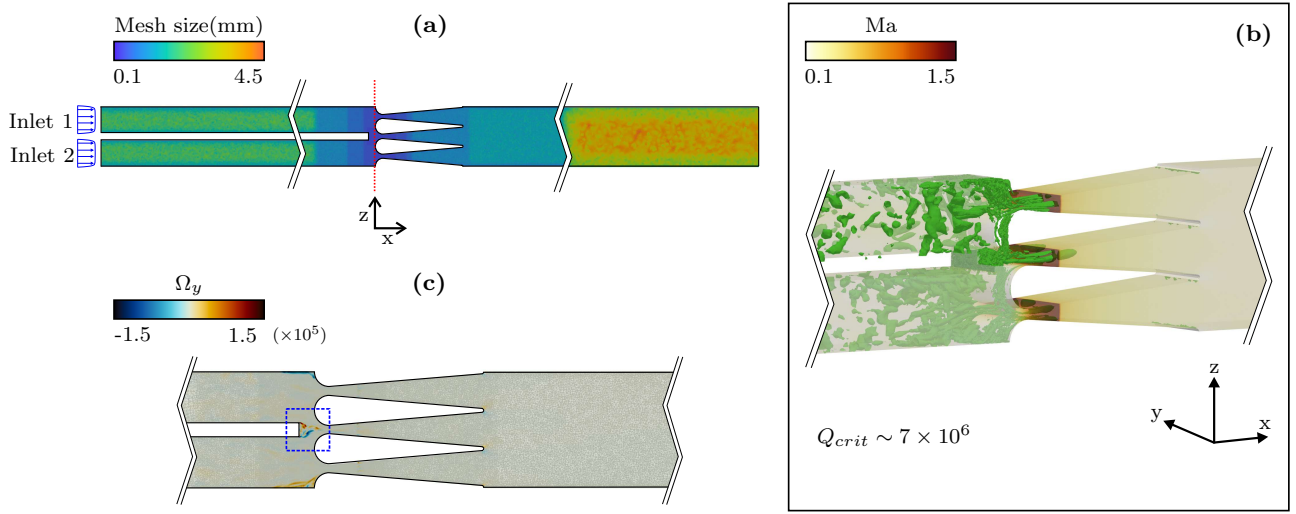}
\end{psfrags}
\caption{Large Eddy Simulations setup of the flow at the crosstalk aperture. (a) Computational domain and mesh size distribution. (b) Instantaneous solution of the simulation showing the vortical structures by means of the $Q$ criterion and the Mach number.  (c) Instantaneous distribution of the $y$ component of the vorticity $\Omega_y$.}
\label{fig:LESdomain} 
\end{figure}

Large Eddy Simulations (LES) of the misaligned vane configuration were performed using the compressible solver AVBP, for three crosstalk aperture lengths $d$ of 3.7, 7.2, and 11.2 mm. For these simulations, a third order fully-explicit two-step Taylor-Galerkin finite element numerical scheme \cite{colin2000} is used. Navier–Stokes Characteristic Boundary
Conditions (NSCBC) are imposed at the inlet and outlet of the domain \cite{poinsot1992}. The relaxation coefficients are tuned to obtain nearly anechoic main inlet and outlet. Boundary
layers are modelled using the classical law-of-the-wall. The computational mesh is generated using Gmsh \cite{Geuzaine2009} and consists of 19 million tetrahedral cells. Figure \ref{fig:LESdomain}a shows the central cut of the fluid domain. The cell sizes are kept coarse at the inlet and outlet channels and finely refined in the crosstalk aperture region and in the contraction where the normal shock is located.  The inlet flows in both channels have a turbulent power-law profile of exponent 0.7. The subgrid Reynolds stress is modelled with the classical Smagorinsky approach \cite{smagorinsky1963} with a typical constant of 0.17. Figure \ref{fig:LESdomain}b  shows a magnified view of the crosstalk aperture and of the turbulent structures in the channel flow upstream of the sonic region between the vanes. 
An instantaneous snapshot of the $y$ component of the total vorticity is presented in \ref{fig:LESdomain}c. The coherent flucturations of the vorticity will be used the next section to compute the ``vortex sound'' radiated for sufficiently large crosstalk aperture length $d$ in the misaligned configuration. 
\begin{figure} 
\centering
\begin{psfrags}
\psfrag{(a)}[][][1.1]{\scriptsize \textbf{(a)}}
\psfrag{(b)}[][][1.1]{\scriptsize \textbf{(b)}}
\psfrag{(ca)}[][][1.1]{\scriptsize \textbf{(c)}}
\psfrag{-1000}[][][1]{\scriptsize -1000}
\psfrag{1000}[][][1]{\scriptsize 1000  (Pa)}
\psfrag{-1500}[][][1]{\scriptsize -1500}
\psfrag{1500}[][][1]{\scriptsize 1500 (Pa)}
\psfrag{p}[][][1]{\scriptsize $\widetilde{p}+{p}^\prime$}
\psfrag{-851.2}[][][1]{\hspace{-10mm}\scriptsize -851}
\psfrag{0.0}[][][1]{~~~\scriptsize 0}
\psfrag{28.5}[][][1]{~~\scriptsize 28.5}
\psfrag{-28.5}[][][1]{~~\scriptsize -28.5}
\psfrag{z}[][][1]{\scriptsize z}
\psfrag{x}[][][1]{\scriptsize x}
\includegraphics[width=17cm]{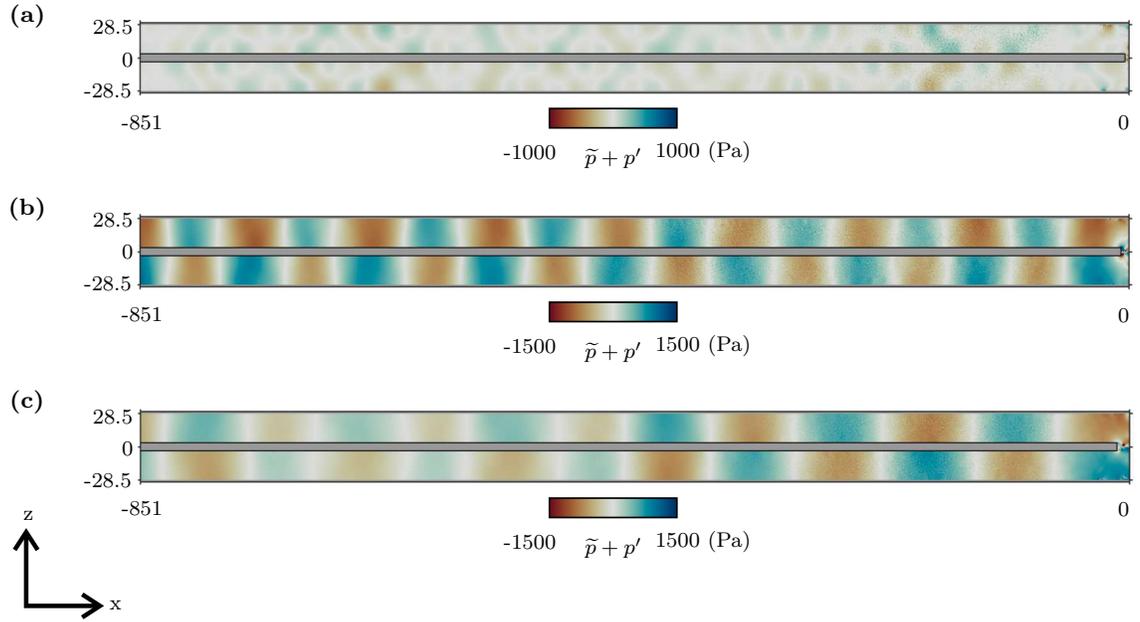}
\end{psfrags}
\caption{Instantaneous  pressure fluctuations obtained by subtracting the mean pressure $\bar{p}$ to the total pressure $p$ in the two channels. The resulting pressure is the sum of the coherent pressure oscillations at the whistling frequency $\tilde{p}$ and the pressure fluctuations associated to the turbulence of the flow $p'$. The top (a), middle (b) and bottom (c) pressure fluctuation fields correspond respectively to $d =3.7$ mm, $d = 7.2$ mm, and $d = 11.2$ mm. The crosstalk aperture is located on the right of the domain at $x=0$.}
\label{fig:LESsolutions} 
\end{figure}
\subsection{Simulations of the intrinsic aeroacoustic instability}
Three configurations were simulated, with $d =3.7$ mm, $d = 7.2$ mm, and $d = 11.2$ mm. Instantaneous snapshots of the pressure fluctuations are presented in Fig.  \ref{fig:LESsolutions}. One can see that, while the flow in the case of the smallest crosstalk aperture does not lead plane acoustic waves in the channels  (Fig.  \ref{fig:LESsolutions}a), the flow in the other two cases leads to an intense whistling with high amplitude plane waves propagating  upstream in the two channels (see Figs.  \ref{fig:LESsolutions}b and   \ref{fig:LESsolutions}c). In the latter cases, the tonal noise radiation from the crosstalk aperture occurs at different acoustic wavelengths. Indeed, the self-oscillation of the shear flow in the crosstalk aperture has a lower oscillation frequency for $d=11.2$ mm than for $d=7.2$ mm, with $f\simeq2030$ Hz and $f\simeq2760$ Hz respectively.  This is in good agreement with the experimental data presented in Fig. \ref{fig:mis_all}a, with a relative difference of the peak frequency of less than 10\%. In Figure \ref{fig:mis_all}d, the PDF of the acoustic pressure measured in one of the channel for the aperture length $d=3.7$ mm has a Gaussian like shape (brown curve), which corresponds to a linearly stable aeroacoustic system, while the one for the aperture length of 11.2 mm has a bimodal distribution (dark green curve), which corresponds to self-oscillations. This key difference is well reproduced by the LES. For the intermediate case of $d=7.2$, the PDF from the experiments  (yellow curve) is Gaussian-like, which indicates that the system is stable, and which is not in agreement with the self-oscillations obtained in the LES. However, this disagreement can be explained by the proximity of the instability threshold when $d\approx 8$ mm and the corresponding high sensitivity of the whistling phenomenon to small geometrical imperfection which may not be considered in the LES. Indeed, one can see in Fig. \ref{fig:mis_all}a that  increasing the aperture length $d$ to 8.7 mm (light green curve) leads to a sharper peak in the PSD and to a flattened unimodal PDF corresponding to a marginally stable aeroacoustic system. 
Finally, an interesting observation from the self-oscillations obtained in the LES of the geometries with $d=7.2$ and $d=11.2$ mm is that the amplitude of the radiated waves exhibit noticeable fluctuations around the mean whistling amplitude. These fluctuations are more pronounced in the case of $d=11.2$ mm for which we see that the waves reaching the anechoic termination on the left side of the computational domain have, for this particular instantaneous snapshot, an amplitude which is noticeably lower than the one of the waves in the right half of the simulated domain. It will be shown in the next section that these fluctuations are associated with the turbulence induced perturbations of the coherent self-oscillations of the shear layers in the crosstalk aperture. 
\subsection{Spatio-temporal evolution of the Lamb vector}
In this section, we investigate the spatio-temporal evolution of the fluctuating component of the Lamb vector $\boldsymbol{\omega}\times \boldsymbol{u}$, which governs the production and dissipation of sound at the crosstalk aperture. Indeed, according to Howe’s energy corollary \cite{howe1980dissipation}, the projection of this vector onto the acoustic velocity field is a source term in the local balance of acoustic energy. 
\begin{figure} 
\centering
\begin{psfrags}
\psfrag{(a)}[][][1.1]{\scriptsize \textbf{(a)}}
\psfrag{(b)}[][][1.1]{\scriptsize \textbf{(b)}}
\psfrag{bin3}[][][1.1]{\scriptsize \textbf{$\phi=60^{\circ}$}}
\psfrag{bin5}[][][1.1]{\scriptsize \textbf{$\phi=120^{\circ}$}}
\psfrag{bin7}[][][1.1]{\scriptsize \textbf{$\phi=180^{\circ}$}}
\psfrag{bin9}[][][1.1]{\scriptsize \textbf{$\phi=240^{\circ}$}}
\psfrag{binII}[][][1.1]{\scriptsize \textbf{$\phi=300^{\circ}$}}
\psfrag{W}[][][1]{~~~\scriptsize $\overline{u}_{z}+\widetilde{u}_{z}$}
\psfrag{w}[][][1]{\scriptsize $\widetilde{u}_{z}$}
\psfrag{-70}[][][1]{\scriptsize $-70$}
\psfrag{70}[][][1]{\scriptsize 70}
\psfrag{-6}[][][1]{\scriptsize $-6$}
\psfrag{6}[][][1]{\scriptsize 6}
\psfrag{-8}[][][1]{\scriptsize $-8$}
\psfrag{8}[][][1]{\scriptsize 8}
\psfrag{-0.01}[][][1]{\scriptsize -10}
\psfrag{0.0}[][][1]{\scriptsize 0}
\psfrag{0.01}[][][1]{\scriptsize 10}
\psfrag{-0.08}[][][1]{\hspace{-3mm}\scriptsize $-17$}
\psfrag{-0.06}[][][1]{~~~\scriptsize 8}
\psfrag{z}[][][1]{\scriptsize $z$}
\psfrag{x}[][][1]{\scriptsize $x$}
\includegraphics[width=16cm]{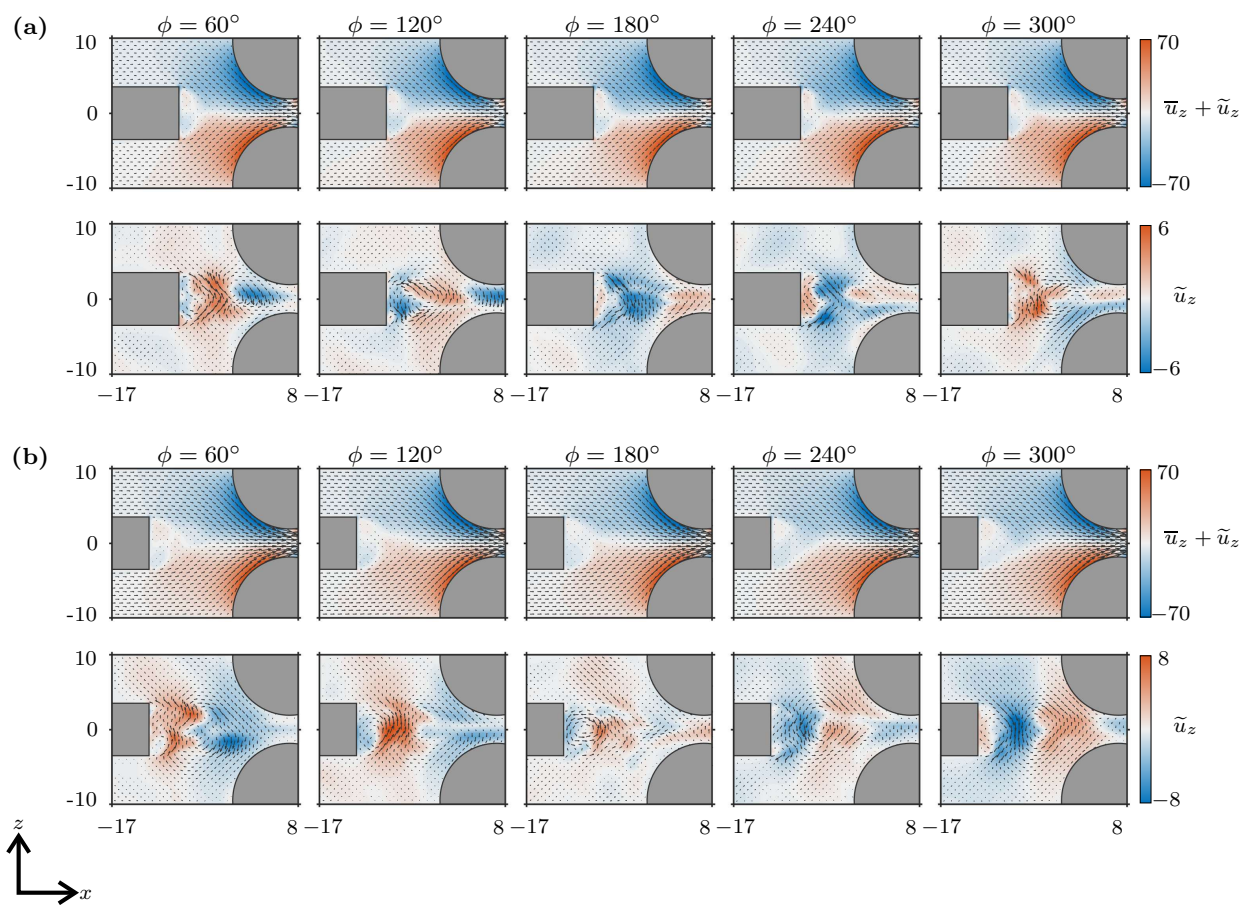}
\end{psfrags}
\caption{Phase-averaged transverse component of the flow velocity  $\overline{u}_{z}+\widetilde{u}_{z}$ and of the coherent velocity fluctuations $\widetilde{u}_{z}$, for the  crosstalk aperture lengths $d=7.2$ mm (a) and $d=11.2$ mm (b) at several phases of the self-sustained acoustic oscillations. The corresponding phase-averaged vector fields are also shown with the arrows.}
\label{fig:PhaseAvFlow}
\end{figure}
\noindent We focus on the cases of $d=7.2$ and $d=11.2$ mm for which the compressible LES yield self-sustained aeroacoustic oscillations. In Fig. \ref{fig:PhaseAvFlow}, the phase-averaged  transverse component of the velocity field $\overline{u}_{z}+\widetilde{u}_{z}$ at the aperture is presented for the two aperture lengths. One can make the following conclusions from these sequences: First, the self-oscillations of the flow are not easy to perceive from  $\overline{u}_{z}+\widetilde{u}_{z}$ and only becomes visible when the mean velocity $\overline{u}_{z}$ is subtracted. Focusing on the  coherent fluctuations $\widetilde{u}_{z}$ whose magnitude are about ten times smaller than the transverse component of the mean flow velocity, one can see an alternance of upward (red) and downward (red) flow deflection, with an advection of the oscillating pattern by the mean flow. Second, one can also see that the convective wavelength is of the order of the distance between the upstream edge of the aperture and the central contraction of the  vanes. It indicates that the aeroacoustics of this flow configuration leads to self-oscillation of the first hydrodynamic mode in the aperture. \\
It is then interesting to compute at each instant the spatial distribution of the fluctuations of the vertical ($z$) component of the Lamb vector for the three crosstalk aperture lengths $d$. 
This post-processing has been done in \cite{boujo20} with particle image velocimetry data in the case of the whistling of a bottle subject to a grazing flow on its top. In the present quasi-two dimensional configuration, we approximate it as follows:  $f_z\simeq \omega_yu_x-\overline{\omega_yu_x}$.  Then, $f_z$ is integrated at each axial position $x$ along the transverse direction $z$ in order to highlight the wave structure of the fluctuations of the vertical component of the Lamb vector. \\
The results are presented in Fig. \ref{fig:Lamb}.
One can see that for $d=3.7$ mm, the amplitude of $\int f_{z}\,d{z}$ is small, as the corresponding acoustic pressure fluctuations $p-\bar{p}=\tilde{p}+p'$. In contrast, for $d=7.2$ mm and $d=11.2$ mm, $\int f_{z}\,d{z}$ exhibits, as in  Fig. 2a of \cite{boujo20}, a slanted pattern. In Fig. \ref{fig:Lamb}e and \ref{fig:Lamb}f, the angle of the waves in this spatio-temporal evolution of $\int f_{z}\,d{z}$ can be clearly seen and it corresponds to the advection velocity. \begin{figure} 
\centering
\begin{psfrags}
\psfrag{(a)}{\scriptsize \textbf{(a)~~$d=3.7$ mm}}
\psfrag{(b)}{\scriptsize \textbf{(b)~~~~~~~$d=7.2$ mm}}
\psfrag{(g)}{\scriptsize \textbf{(c)~~~~~~~~~~~~~$d=11.2$ mm}}
\psfrag{(d)}{\scriptsize \textbf{(d)}}
\psfrag{(e)}{\scriptsize \textbf{(e)}}
\psfrag{(f)}{\scriptsize \textbf{(f)}}
\psfrag{-5}[][][1]{\scriptsize ~~$-1$}
\psfrag{+5}{\scriptsize $1$ (arb. unit)}
\psfrag{x107}[][][0.8]{~}
\psfrag{10}[][][1]{\scriptsize 10}
\psfrag{0}[][][1]{\scriptsize 0}
\psfrag{+500}[][][1]{\scriptsize 500}
\psfrag{-500}[][][1]{\scriptsize -500}
\psfrag{4}[][][1]{\scriptsize 4}
\psfrag{3.7}[][][1]{\scriptsize 3.7}
\psfrag{7.2}[][][1]{\scriptsize 7.2}
\psfrag{II.2}[][][1]{\scriptsize 11.2}
\psfrag{t(ms)}[][][1]{\scriptsize t(ms)}
\psfrag{p}[][][1]{\scriptsize $p-\bar{p}$}
\psfrag{f}[][][1]{\scriptsize $\int f_{z}\,d{z}$}
\includegraphics[scale=0.75]{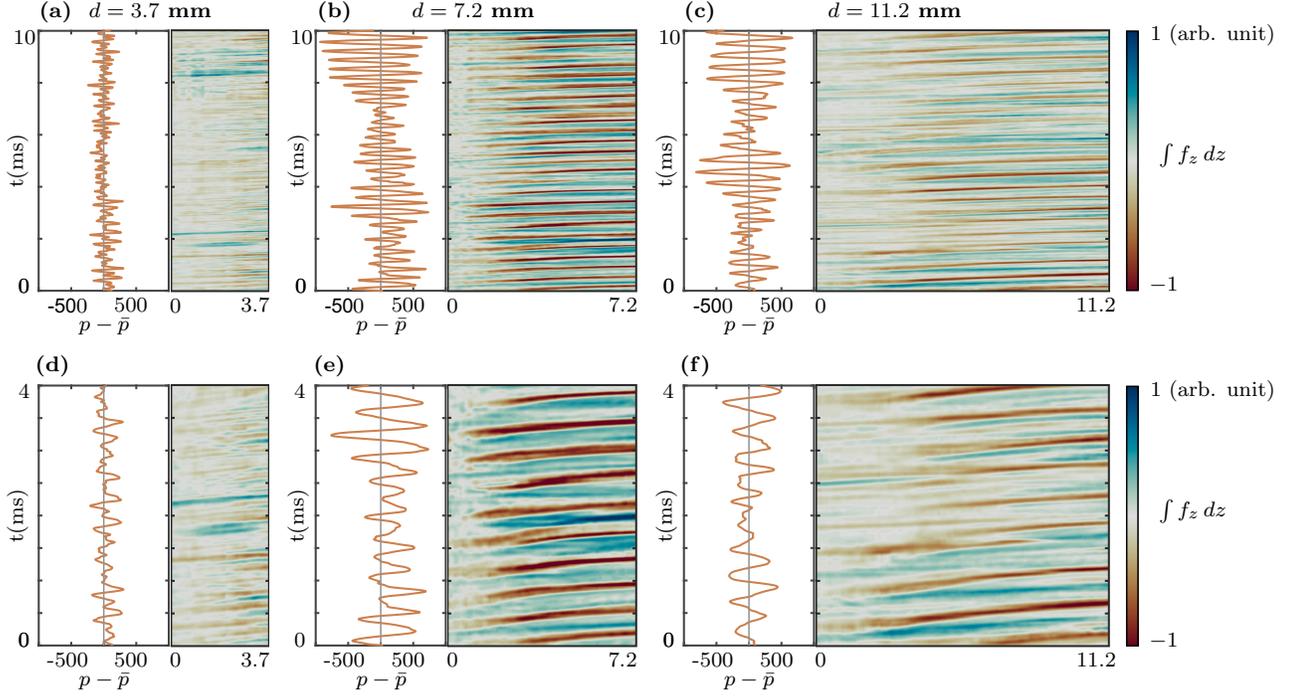}
\end{psfrags}
\caption{Time traces of the pressure fluctuations $p-\bar{p}$ in the channels, and space-time evolution of the vertical component of the fluctuations of the Lamb vector $f_z\simeq \omega_yu_x-\overline{\omega_yu_x}$ integrated vertically at each streamwise location for $d=3.7$ mm (a,d), $d=7.2$ mm (b,e) and $d=11.2$ mm (c,f). The top row (a,b,c) presents the results over a time interval of 10 ms to highlight the fluctuations of the self-oscillation amplitude. The bottom row (d,e,f) presents the results over a time interval of 4 ms to highlight the convective nature of the $\int f_{z}\,d{z}$ in the aperture.}
\label{fig:Lamb}
\end{figure}
Considering the acoustic pressure time traces of Fig. \ref{fig:Lamb}b and \ref{fig:Lamb}c, one  can see the change of dominant oscillation frequency with the aperture length $d$. It is also interesting to see the correlation between the magnitude of the oscillations of $\int f_{z}\,d{z}$ and the ones of the resulting acoustic pressure in the channels. These fluctuations of the vortex sound amplitude originate from the intense turbulence in the channels, which can be visualised in Fig. \ref{fig:LESdomain}b, and which perturbs the  intrinsic aeroacoustic self-oscillations in the aperture.
\section{Conclusion}
Intrinsic aeroacoustic instabilities at a crosstalk aperture between two channels  upstream of choked vanes was observed and investigated for the first time experimentally and with Large Eddy Simulations (LES). Different geometries of the crosstalk aperture were considered, with variations of its area, of the shape of its upstream edge and of its alignment with the row of choked vanes. The Strouhal and Helmholtz numbers of this generic configuration are relevant for the crosstalk aperture between neighbouring combustion chambers in heavy duty gas turbines featuring can-annular combustors.  The whistling phenomenon is stronger for an aperture with rectangular upstream edge, which is not aligned with a central vane. The upstream anechoic condition of the  channels eliminates the possibility of  longitudinal modes having a feedback action on the aeroacoustic dynamics at the aperture. This configuration therefore implies that the whistling results from intrinsic aeroacoustic modes. The Strouhal scaling enables prediction of the of the peak frequency of the self-oscillations. The probability density function of the acoustic pressure measured in the channels is used to demonstrate a bifurcation from linearly-stable aeroacoustic oscillations to self-sustained ones. The intrinsic aeroacoustic instability is reproduced with compressible LES. The simulations of three of the experimentally studied configurations allows us to unravel the oscillation flow upstream of the choked vanes and to identify the source of acoustic energy associated to the coherent oscillation of the shear layer in the crosstalk aperture. These results are crucial for industrial configuration, because we show that even without acoustic feedback from upstream boundary, such crosstalk aperture can self-oscillate at very large acoustic amplitudes. We can therefore surmise that the reflection of sound from the burner and the flames in a real can-combustor configuration can further amplify this aeroacoustic self-oscillations, resulting in a potentially  detrimental aero-thermo-acoustic limit cycle.

\section*{Acknowledgements}
Financial support from the ETH Foundation, under the recommendation of GE Vernova, is gratefully acknowledged.

\section*{Authorship contribution statement}
\textbf{A.B.} performed the experiments, analyzed the experimental data and generated the figures. \textbf{K.P.} performed the simulations, analyzed the numerical data and generated the figures. \textbf{N.N.} conceived the research idea, designed the test rig and supervised the project. \textbf{N.N.} and \textbf{B.S.} designed the vanes and guided the investigations. All authors discussed the results. \textbf{A.B.} and \textbf{N.N.}  wrote the manuscript.

\section*{Declaration of competing interest}
The authors declare that they have no known competing financial interests or personal relationships that could have appeared to influence the work reported in this paper.

\begin{figure} 
\centering
\begin{psfrags}
\psfrag{0}[][][1.1]{\scriptsize 0}
\psfrag{80}[][][1.1]{\scriptsize 80}
\psfrag{100}[][][1.1]{\scriptsize 100}
\psfrag{120}[][][1.1]{\scriptsize 120}
\psfrag{140}[][][1.1]{\scriptsize 140}
\psfrag{70}[][][1.1]{\scriptsize }
\psfrag{90}[][][1.1]{\scriptsize }
\psfrag{110}[][][1.1]{\scriptsize }
\psfrag{130}[][][1.1]{\scriptsize }
\psfrag{150}[][][1.1]{\scriptsize }
\psfrag{1000}[][][1.1]{\scriptsize }
\psfrag{2000}[][][1.1]{\scriptsize 2000}
\psfrag{3000}[][][1.1]{\scriptsize }
\psfrag{4000}[][][1.1]{\scriptsize 4000}
\psfrag{5000}[][][1.1]{\scriptsize }
\psfrag{6000}[][][1.1]{\scriptsize 6000}
\psfrag{PSD}[][][1.1]{\scriptsize SPL [dB]}
\psfrag{freq}[][][1.1]{\scriptsize Freq. [Hz]}
\psfrag{mdot}[][][1.1]{\scriptsize  $\dot{m}_{\text{air}}$ [g/s]}
\psfrag{massflow1}[][][1.1]{\scriptsize 80 g/s}
\psfrag{massflow2}[][][1.1]{\scriptsize 90 g/s}
\psfrag{massflow3}[][][1.1]{\scriptsize 100 g/s}
\psfrag{massflow5}[][][1.1]{\scriptsize 110 g/s}
\psfrag{massflow6}[][][1.1]{\scriptsize 120 g/s}
\psfrag{a}[][][1.1]{\scriptsize a)}
\psfrag{b}[][][1.1]{\scriptsize b)}
\psfrag{d7}[][][1.1]{\scriptsize 11.2}
\includegraphics[width=17cm]{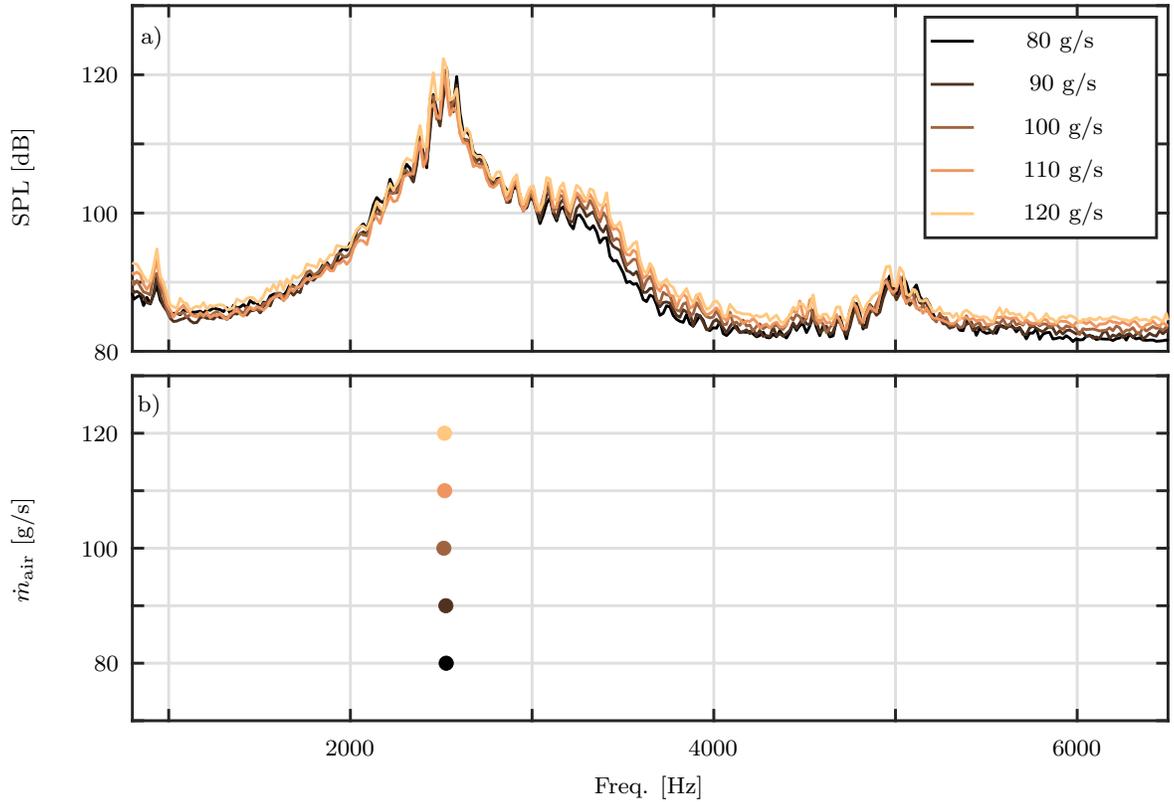}
\end{psfrags}
\caption{a) Acoustic PSD for five different mass flow rates  $\dot{m}_{\text{air}}$. The crosstalk aperture geometry is with $d=7.2$ mm and with the rectangular edge. For all mass flow rates, the flow through the vanes is choked. b) Frequency of the dominant peak for the five different mass flow rates. }
\label{fig:Spec_mdot} 
\end{figure}

\appendix
\section*{Appendix}
As mentioned in Section \ref{sec:expSetup}, all the results presented in the main body of the article correspond to the same mass flow. Once the vanes are choked, an increase in the  mass flow rate is accompanied with an of the mean pressure and density while the flow velocity does not vary. 
Figure \ref{fig:Spec_mdot}a shows the acoustic PSD  for the $d=7.2$ mm aperture with rectangular upstream edge and for different total mass flow rates beyond the choked flow condition, between 80 and 120 g/s. All the spectra are superimposed with each others, which shows that the Strouhal scaling can be used for any operating pressure. Figure \ref{fig:Spec_mdot}b shows the frequency of the dominant peak as function of the air mass flow rate $\dot{m}_{air}$.

\bibliographystyle{elsarticle-num.bst}
\bibliography{references}

\end{document}